\documentstyle[11pt,preprint]{aastex}

\newcommand{\be}{\begin{equation}}
\newcommand{\ee}{\end{equation}}

\newcommand{\kms}{km \hskip -2pt s$^{-1}$}
\newcommand{\mkms}{{\rm km}\,{\rm s}^{-1}}
\newcommand{\hmpc}{\;h^{-1}\;{\rm Mpc}}
\newcommand{\hubunits}{\mkms\;{\rm Mpc}^{-1}}

\newcommand{\Om}{\Omega_{m,0}}       
\newcommand{\Or}{\Omega_{r,0}}       
\newcommand{\Op}{\Omega_{\phi,0}}    
\newcommand{\Oi}{\Omega_{i,0}}       
\newcommand{\Omz}{\Omega_{m}(z)}       
\newcommand{\Orz}{\Omega_{r}(z)}       
\newcommand{\Opz}{\Omega_\phi(z)}    

\newcommand{\lya}{{Ly$\alpha$}}

\def\la{\mathrel{\spose{\lower 3pt\hbox{$\mathchar"218$}}
     \raise 2.0pt\hbox{$\mathchar"13C$}}}
\def\ga{\mathrel{\spose{\lower 3pt\hbox{$\mathchar"218$}}
     \raise 2.0pt\hbox{$\mathchar"13E$}}}

\newbox\grsign \setbox\grsign=\hbox{$>$} \newdimen\grdimen \grdimen=\ht\grsign
\newbox\simlessbox \newbox\simgreatbox
\setbox\simgreatbox=\hbox{\raise.5ex\hbox{$>$}\llap
     {\lower.5ex\hbox{$\sim$}}}\ht1=\grdimen\dp1=0pt
\setbox\simlessbox=\hbox{\raise.5ex\hbox{$<$}\llap
     {\lower.5ex\hbox{$\sim$}}}\ht2=\grdimen\dp2=0pt
\def\spose#1{\hbox to 0pt{#1\hss}}

\begin{document}

\title{Prospects For Determining The Equation Of State Of The
Dark Energy: What Can Be Learned From Multiple Observables?}

\author{Jens Kujat}
\affil{Department of Physics, Ohio State University, Columbus, OH  43210}

\author{Angela M. Linn}
\affil{Department of Physics, Ohio State University, Columbus, OH  43210}

\author{Robert J. Scherrer}
\affil{Department of Physics and Department of Astronomy,
Ohio State University, Columbus, OH 43210}

\author{David H. Weinberg}
\affil{Department of Astronomy, Ohio State University, Columbus, OH 43210}

\begin{abstract}
The dark energy that appears to produce the accelerating expansion of the universe
can be characterized by an equation of state
$p=w\rho$ with $w<-1/3$.
A number of observational tests have been proposed to study the value or
redshift dependence of $w$,
including SN Ia distances, the Sunyaev-Zel'dovich
effect, 
cluster abundances, strong and weak gravitational lensing, galaxy and quasar 
clustering,
galaxy ages, the \lya\ forest, and cosmic microwave background anisotropies.
The proposed observational tests based on these phenomena measure either the
distance-redshift relation $d(z)$, the Hubble parameter $H(z)$, the age of the
universe $t(z)$, the linear growth factor $D_1(z)$, or some combination of these
quantities. We compute the evolution of these four observables, and of the 
combination $H(z)d(z)$ that enters the Alcock-Paczyznski anisotropy test, in 
models with constant $w$, in quintessence models with some simple forms
of the potential $V(\phi)$, and in toy models that allow more radical time
variations of $w$. Measurement of any of these quantities to precision of a few
percent is generally sufficient to discriminate between $w=-1$ and $w=-2/3$.
However, the time-dependence predicted in quintessence models is extremely
difficult to discern because the quintessence component is dynamically 
unimportant at the redshifts where $w$ departs substantially from its low-$z$
value. Even for the toy models that allow substantial changes in $w$ at low
redshift, there is always a constant-$w$ model that produces very similar 
evolution of all of the observables simultaneously. We conclude that measurement
of the effective equation of state of the dark energy may be achieved by several
independent routes in the next few years, but that detecting time-variation in
this equation of state will prove very difficult except in specialized
cases.
\end{abstract}


\vfill
\eject

\section{Introduction}
\label{sec:intro}

The big cosmological surprise of recent years is that the dominant form of
energy in the universe has negative pressure and is therefore causing the
expansion of the universe to accelerate.  The most direct evidence for
acceleration comes from the Hubble diagram of Type Ia supernovae (SN Ia),
in particular the relative apparent brightness of SN Ia at redshifts
$z\sim 0$ and $z\sim 0.5-1$ \citep{riess98,perlmutter99}.  However, 
other strong arguments for a ``dark energy'' component follow from 
combining the cosmic microwave background (CMB) evidence for a spatially
flat universe \citep{netterfield01,pryke01} with either a minimum age 
$t_0 \sim 13\;$Gyr \citep{vandenberg96} or dynamical evidence that the density
of clustered matter is well below the critical density
(see \citealt{bahcall97,carlberg97,weinberg99b} for examples of three
distinct routes to this conclusion, though there are many others).
The first combination, together with a Hubble constant
$H_0 \approx 70\;\hubunits = (14\;{\rm Gyr})^{-1}$
\citep{freedman01}, requires a component whose gravitational acceleration
roughly cancels the gravitational deceleration caused by the 
pressureless matter, so that $t_0 \approx H_0^{-1}$.  The second
combination requires that the dominant form of energy be unclustered,
though it implies nothing more specific about its equation of state.
A more model-dependent argument for a negative pressure component
comes from the success of inflationary models with cold dark matter (CDM)
and a cosmological constant ($\Lambda$) in matching a variety of
constraints from CMB anisotropies and large scale structure measurements
(see \citealt{wang01x} for a recent review).

In this paper, we explore the prospects for determining the equation of
state of the dark energy component through a variety of observational
methods.  A true cosmological constant can be treated
as a vacuum energy with time-independent density and pressure related
by $p=-\rho$.  Current observations favor an equation of state fairly
close to this prediction \citep{garnavich98}.  However, a number of authors
have considered the more general possibility that the negative pressure
component is a scalar field (a.k.a.\ ``quintessence'') with energy
density determined by its potential and effective equation of state 
$p=w\rho$, where $w$ can be constant or time-varying
\citep{ratra88,turner97,caldwell98}.  Interest in models with 
time-varying $w$ has been spurred by arguments that certain simple
potentials lead ``naturally'' to a negative pressure quintessence component 
that dominates the expansion at late times, 
independent of the initial conditions \citep{zlatev99,steinhardt99}.
Variants on this theme include fields with a non-standard kinetic term \citep{kessence}
or models with a complex scalar field \citep{spintessence}.

Further afield, there is the possibility that the negative pressure
component is a network of frustrated topological defects
\citep{vilenkin85,spergel97}, 
or that cosmic acceleration arises from a breakdown
of general relativity rather than the addition of a new energy
component (\citealt{mannheim01}; see also \citealt{tegmark01}).
The hope, thus far unrealized, is that one of these ideas will
eventually provide a natural explanation of why the vacuum energy
density is 120 orders-of-magnitude below the Planck scale and why
it is comparable to the matter density at the present day,
without having to resort to anthropic selection arguments
\citep{efstathiou95,martel98}.

Any clear evidence that $w \neq -1$, or, better still, that $w$
varies in time, would provide crucial clues towards understanding
the physics of the dark energy.  Through its influence
on the cosmic expansion history, this component affects many observable
phenomena, including CMB anisotropies, the \lya\ forest, strong and
weak gravitational lensing, the anisotropy of quasar and galaxy clustering in
redshift space, the ages of the oldest galaxies as a function of
redshift, and standard-candle or standard-ruler measurements of the
distance-redshift relation.  This paper discusses these potential
observational tests in a unified fashion.  The equation of state
determines the history of the energy density $\rho_\phi$, which,
together with the densities $\rho_m$ and $\rho_r$ of matter and
radiation, determines the evolution of the Hubble parameter $H(z)$
via the Friedmann equation.  The history of $H(z)$ in turn determines
the age of the universe $t(z)$, the growth factor of linear perturbations
$D_1(z)$, and distance measures like the angular diameter distance
$d_A(z)$ or luminosity distance $d_L(z)$, which are related to each
other by cosmology-independent powers of $(1+z)$.  Essentially all
proposed tests of the properties of the negative pressure component
amount to measurements of $H(z)$, $t(z)$, $D_1(z)$, or $d(z)$,
or some combination of them, at redshifts accessible to a particular
observational technique.  We will investigate the dependence of these
four quantities, and of the specific combination $H(z)d_A(z)$ that is
constrained by the Alcock-Paczynski (\citeyear{alcock79}; hereafter AP) 
anisotropy test, on the value and time history of $w$.

Our paper joins, and, we hope, complements, a flood of recent papers
that examine the prospects for specific tests and specific data sets
in much greater detail.  Since the strongest evidence for $\Lambda$
or a quintessence component comes from SN Ia observations, and
substantial improvements are likely from ground-based campaigns
and possibly a dedicated satellite (SNAP; see {\tt http://snap.lbl.gov}),
many authors have examined the extent to which present or future
SN Ia observations can constrain $w(z)$
\citep{turner97,garnavich98,astier00,chiba00,hutererturner01,saini00,
barger01,chevallier01,maor01,ng01,podariu01,wang01a,wang01b,weller01}.
Because CMB anisotropy predictions depend most strongly on the
sum of $\rho_\phi$ and $\rho_m$ while SN Ia distances depend more
nearly on the difference, the combination of these complementary observations
yields much tighter constraints on the negative pressure component 
than either does alone \citep{caldwell98,efstathiou99,baccigalupi01,
corasiniti01,doran01}.  The Sunyaev-Zel'dovich 
effect or size of radio sources offer alternative ways of measuring $d_A(z)$
\citep{birkinshaw99,lima01}, and the volume-redshift test using galaxy counts
constrains the combination $d_A^2(z)H^{-1}(z)$ \citep{newman00,newman01}.
The evolution of the galaxy cluster mass function can constrain
the linear growth factor $D_1(z)$ \citep{benabed01,doran01b,haiman01,
newman01b,weller01b},
and population synthesis modeling of galaxy spectra can constrain
$t(z)$ \citep{lima00}.  Jimenez \& Loeb (2001) suggest that 
relative galaxy ages can be used to measure $dz/dt$, and thus $H(z)$.
\cite{hui99a}  and \cite{huterer01}
have examined constraints on $w$
that can be obtained from weak lensing, while \cite{hu}
has considered lensing in combination with the
CMB.
\cite{calvao01} have
discussed constraints that could be obtained by applying the AP
test to the 2dF quasar redshift survey of 
Boyle et al. (\citeyear{boyle00}; for related discussions see
\citealt{hui99,cappi01,dalal01,mcdonald01}).

Most of these papers have considered the potential observational
constraints singly, or in pairs.
The goals of our more abstract discussion, where
we consider all of these observables together but do not focus
on specific observational strategies, are twofold.  First, we aim
to understand what level of precision is necessary with any of
these quantities to obtain useful constraints on $w$.
Second, we want to know whether these different observables
provide complementary information about the time-variation of $w$,
breaking degeneracies that exist for a single measure by probing
different aspects of the expansion history.  Unfortunately,
our conclusions on the latter point are pessimistic --- there
are many different ways to measure $w$, but distinguishing a
time-varying $w$ from a constant $w$ is likely to prove
difficult.  The papers by \cite{wang00} and \cite{tegmark01} also
consider multiple observables, focusing on present
constraints and future prospects, respectively.  
Tegmark's paper, in particular,
is similar in spirit to ours, but different in the way that it
frames the problem and evaluates the prospects.

In the next section we discuss the various quintessence models that
we examine in this paper.  We discuss the observables in 
\S\ref{sec:observables}, beginning with the formulas that relate
these quantities to the expansion history and proceeding to a
brief account of observations that might measure these quantities in
the next few years.  We present our results
in \S\ref{sec:results},
first for the quintessence models described in \S\ref{sec:quintessence},
then for a class of ``toy'' models designed to allow stronger time-variation
of $w$ at low redshift.  We summarize our conclusions 
in \S\ref{sec:conclusions}.

\section{Quintessence Models}
\label{sec:quintessence}

We will adopt the language and calculational framework of quintessence
models, though most of our general conclusions are also relevant to
other possible explanations of cosmic acceleration, like those 
mentioned in \S\ref{sec:intro}.  Also, in light of evidence from
the location of the first acoustic peak in the CMB anisotropy
spectrum \citep{netterfield01,pryke01}, we will restrict our attention
to spatially flat models.

The Friedmann equation for a spatially flat, expanding universe can be written 
\begin{equation}
{\dot{a} \over a} \equiv H(z) = H_0 \sqrt{\sum_i\Oi {\rho_i(z)\over\rho_{i,0}}}
  ~.
\label{eqn:fr}
\end{equation}
Here $a$ is the scale factor, $\dot{a}$ is the derivative of the scale factor
with respect to time $t$, $H_0$ is the value of the Hubble parameter at the
present time $t_0$, and $\Oi$ is the present density of some 
$i$th component of the
energy density relative to the present critical density 
($\Oi \equiv \rho_{i,0} / \rho_{c,0}$).  
For adiabatic expansion, the energy density of a component with equation
of state $p_i = w_i\rho_i$ with constant $w_i$ evolves with redshift as
\begin{eqnarray}
{\rho_i(z) \over \rho_{i,0}} &=& (1+z)^{n_i} ~, \label{eqn:rhon}\\
n_i &\equiv & 3(1+w_i).  \label{eqn:ndef}
\end{eqnarray}
Normal matter has $w_i=0$ and $n_i=3$, while radiation has $w_i=1/3$
and $n_i=4$.  A true cosmological constant, with $\rho_i=$const, $n_i=0$,
has $w_i=-1$.  We will often refer to models in terms of the energy
density scaling index $n$, defined by equation~(\ref{eqn:rhon}),
rather than by $w$ itself, since the value of $n$ more directly
captures the impact of a component on the expansion history.

A coasting expansion, in which comoving observers have constant
velocity, has $H(z)\propto (1+z)$.  An accelerated expansion
requires, at a minimum, that the dominant energy component have
$n_i<2$, and thus $w_i < -1/3$.  [More precisely, $\langle w \rangle$,
the density-weighted average value
of $w$, must satisfy $\langle w \rangle < -1/3$].
Quintessence, a term reintroduced
to cosmology by \cite{caldwell98} after millennia of neglect,
refers generically to a scalar field with equation of state 
$p_\phi = w_\phi\rho_\phi$ and $w_\phi<0$.
The first class of models that we consider are those in which
$w_\phi$ is constant.
In this case, the Friedmann equation can be written
\begin{equation}
H(z) \equiv {\dot{a} \over a} = H_0 \sqrt{\Or(1+z)^4 + \Om(1+z)^3 +
\Op(1+z)^{n}} ~,
\label{eqn:hz}
\end{equation}
with $n$ given by equation~(\ref{eqn:ndef}).

More general models often treat quintessence as a minimally coupled scalar
field $\phi$, obeying the equation
\begin{equation}
\label{phiev}
\ddot{\phi}  =  - 3 H \dot{\phi} - \frac{dV}{d\phi} ~,
\end{equation}
where $w$ for the scalar field is
\begin{equation}
w={(1/2)\dot \phi^2 - V(\phi) \over (1/2)\dot \phi^2 + V(\phi)} ~.
\end{equation}
When $V(\phi)$ is an exponential or a negative power-law,
the scalar field has the desirable
property that its final evolution is independent of initial
conditions, a behavior that has been dubbed ``tracking"
\citep{zlatev99,steinhardt99}.
The negative power-law potentials lead to
constant $w$ when the contribution from the scalar field energy
density is sub-dominant
\citep{ratra88,liddle99}, but when the scalar field
energy density comes to
dominate at late times, the value of $w$ changes.  In principle, then,
such models should be observationally distinguishable from models with
constant $w$.

For our second class of models, we have chosen a subset of
the negative power-law potentials, where
\begin{equation}
V(\phi) \propto \phi ^ \alpha,
\end{equation}
with $\alpha < 0$.  If the dominant component has a
density that scales as $\rho \propto (1+z)^m$ (e.g., $m=4$
during the radiation-dominated era and $m=3$ during the matter
dominated era), then these models have
\begin{equation}
\label{eqn:scaling}
n = [\alpha/(\alpha-2)] m
\end{equation}
when $\rho_\phi \ll \rho_m$ \citep{liddle99}.  At late times, when the scalar
field energy density begins to dominate, equation (\ref{eqn:scaling}) no
longer holds, and $n$ changes with time.  We have chosen to examine
two representative cases:  $\alpha = -1$ and $\alpha = -6$.
This choice is somewhat arbitrary, but these are the same cases
that are discussed by \citet{zlatev99}.

The evolution of $n$ in these models is displayed in Figure 1,
assuming cosmological parameter values $\Om=0.4$, $\Or=9.8\times 10^{-5}$,
and $\Op=1-\Om-\Or$.  (This value of $\Or$ corresponds to a
photon temperature of $T = 2.73$ K, a standard neutrino population,
and a Hubble parameter of $H_0 = 65\;\hubunits$.  This is the only
place in our calculations where $H_0$ enters, and it has a very
small effect on our results).
For these models, we define $n(z)$ to be the local logarithmic
derivative of $\rho_\phi$ with respect to $(1+z)$.
Figure~1 shows that the value of $n$ for $3 \la z \la 10$ is
almost exactly constant and given by equation~(\ref{eqn:scaling}), namely
$n = 1$ for $\alpha = -1$ and $n = 9/4$ for $\alpha = -6$.
At $z < 3$, $n$ decreases slightly, reaching present-day
values of $n = 0.77$ for $\alpha = -1$ and $n = 1.89$ for $\alpha = -6$.
In \S\ref{sec:results} we will see whether cosmological
tests can detect these slight changes in $n$.

\begin{figure}[tbp]
   \label{Fig:n_vs_z}
   \centering
   \includegraphics{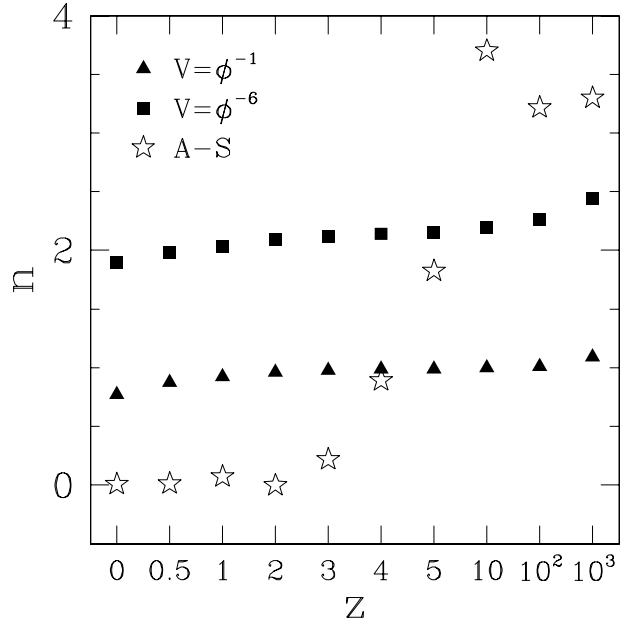}
   \caption{The evolution of $n \equiv d\log \rho_\phi / d\log (1+z)$
   as a function of redshift $z$, for a scalar field with
   the indicated potential and a cosmological model with
   $\Om=0.4$, $\Or=9.8\times 10^{-5}$, and $\Op=1-\Om-\Or \approx 0.6$.
   In this figure, and in all other
   figures in the paper, the horizontal axis effectively represents a set of
   discrete bins, so it is neither linear
   nor logarithmic.  
   }
\end{figure}

Although models with an exponential potential are quite natural
and can lead to $\rho_\phi \approx \rho_m$ at all times,
they are ruled out for several reasons:  they tend to
give $\rho_\phi$ too large during primordial nucleosynthesis
\citep{bbnref}, and they lead to $n = 3$ at late times, which
does not produce an accelerated expansion.
These problems are remedied in the model of
\citet{albrecht00}, who introduced a potential consisting of an
exponential multiplied by a polynomial,
\begin{equation}
V(\phi) = [ ( \phi - B ) ^ \gamma + A ]\, e^{-\lambda \phi},
\end{equation}
where $A$, $B$, $\gamma$, and $\lambda$ are constants.
These constants can be chosen to produce a model for which
$n = m$ at early times, when the scalar field is sliding down
the exponential potential, but
$n=0$ at late times, when the scalar field settles
into the local minimum in the potential.
The constants in this model must still be tuned to give
the desired value for $\Op$;
following Albrecht and Skordis (2000), we have examined a model with $A=0.01, B=34.8,
\gamma=2,$ and $\lambda=8$, and with initial conditions chosen so as to fix $\Om
= 0.4$ today.  The evolution of $n$ for this model is also shown
in Figure 1.  It exhibits a sharp transition from 
$n\approx 3.5$ at $z \geq 10$ to $n\approx 0$ at $z<3$, a much more dramatic
change than that in the power-law scalar field models.
[Exponential potentials can also be made to work in models
in which the scalar field is coupled to matter \citep{amendola}; however,
we confine our attention in this paper to minimally-coupled fields].

\begin{figure}[tbp]
   \label{Fig:Omega_vs_z}
   \centering
   \includegraphics{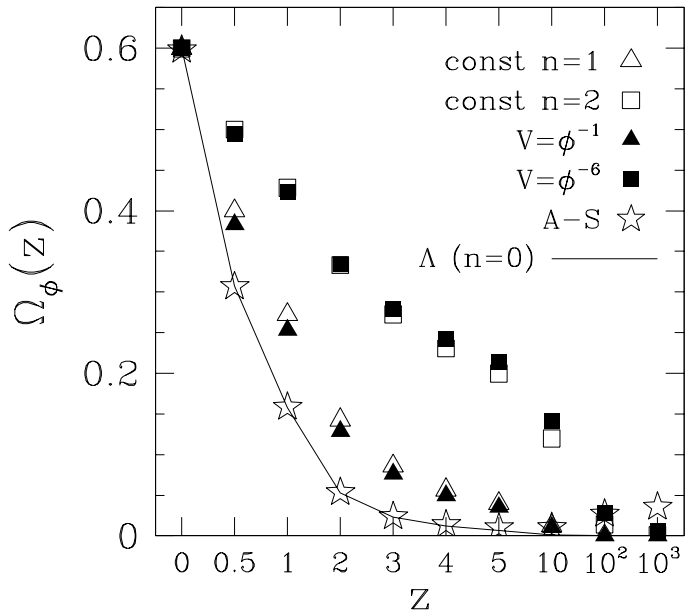}
   \caption{$\Omega_\phi$ as a function of redshift $z$ for the indicated
   quintessence models.  The solid curve is $\Omega_\Lambda(z)$
   for a cosmological constant. }
\end{figure}

The dynamical significance of the quintessence component is quantified
by the density parameter $\Opz$.  Figure~\ref{Fig:Omega_vs_z} shows the
evolution of $\Opz$ for a cosmological constant (solid curve) and the
five quintessence models discussed above.  The $n=0$ (cosmological
constant), $n=1$, and $n=2$ cases are quite distinct, as one would
expect from their differing $\rho_\phi(z)$.  However, 
the $V(\phi)=\phi^{-1}$ case closely parallels the constant $n=1$ case,  
and the $V(\phi)=\phi^{-6}$ case likewise tracks the model with constant $n=2$.
The Albrecht-Skordis $\Omega_\phi$ is nearly
indistinguishable from that of a cosmological constant except at high
redshift, where the change in $n$ makes a small but noticeable difference.

\section{The Observables}
\label{sec:observables}

Our starting point is the Friedmann equation in the form of
equation~(\ref{eqn:fr}).  As components we consider matter
with $\Om=0.4$, radiation with $\Or=9.8\times 10^{-5}$, and
quintessence with $\Op=1-\Om-\Or\approx 0.6$.
We compute the ratio $\rho_\phi(z)/\rho_{\phi,0}$ from 
equation~(\ref{eqn:rhon}) for constant-$w$ models 
[thus obtaining equation~(\ref{eqn:hz}) for $H(z)$]
or by computing the evolution of $\phi$ 
from the dynamical equation~(\ref{phiev}) for the negative power-law
or Albrecht-Skordis models.

The Friedmann equation directly determines the behavior of
our first observable, the Hubble parameter $H(z)$.
We compute other observables given $H(z)$ via the standard treatments
in, e.g., \cite{peebles80,peebles93}, \cite{kolb90}, or \cite{hogg99}.
The age of the universe at redshift $z$ is
\be
\label{eqn:age}
t(z)= \int_z^\infty \frac{dz'}{(1+z') H(z')} ~.
\ee
The angular diameter distance $d_{A}(z)$, which is the 
ratio of the comoving size of an object to its angular size in radians, is 
\be
\label{eqn:da}
d_{A}(z)= {c \over (1+z)} \int_0^z \frac{dz'}{H(z')} ~.
\ee
Other distances, e.g., those that affect the SN Ia Hubble diagram or
gravitational lensing predictions, are related to $d_A$ by powers of
$(1+z)$; the bolometric luminosity distance, for example, is
$d_L(z) = (1+z)^2 d_A(z)$.  
Since these factors are independent of the cosmological model, a measurement
of any of these distances determines all of them to the same fractional
accuracy, so we take $d_A(z)$ as our representative observable for all 
distance measures.

The linear growth factor $D_1$ is defined by the relation
\begin{equation}
\delta^{(1)}({\bf x},t) = \delta({\bf x}) D_1(t),
\end{equation}
where $\delta^{(1)}({\bf x},t)$ is the first-order density perturbation.
We choose the normalization $D_1(z=0)=1$,
so that $D_1(z) = \delta^{(1)}(z)/\delta^{(1)}(0)$ gives the linear growth of
perturbations
between redshift $z$ and redshift 0.  Then $D_1$
is the growing-mode solution to the
differential equation
\begin{equation}
\label{evol}
\ddot{D}_1 + 2 H(z)\dot{D}_1 - {3\over2}\Om H_0^2 (1+z)^3 D_1 = 0.  
\label{eqn:D1}
\end{equation}
For fixed $\Om$, $D_1(z)$ is a function only of $H(z)$,
so it is again determined by the Friedmann equation.  
We solve this equation for $D_1(z)$ with a standard Runge-Kutta
integration method.  
In the pure cosmological constant case, a closed form expression for
$D_1(z)$ is 
\be
D_1(z) = {H(z)\over H_0} 
         \int_{z}^\infty {dz' (1+z')\over H^3(z')}
         \left[\int_0^\infty {dz' (1+z')\over H^3(z')}\right]^{-1} ~,
\label{eqn:D1int}
\ee
where the factor in brackets enforces our normalization definition
(\citealt{eisenstein97}, based on \citealt{heath77}).
Unfortunately, this expression generalizes only
to the case $n=2$, but it does illustrate that the
linear growth factor weights the expansion history in a different
way than does the age or distance.  For constant $w$,
a solution for $D_1(z)$ can be found in terms of hypergeometric
functions (Silveira \& Waga 1994).

In addition to these four observables, we consider the specific 
combination $h(z)$ that is probed by the AP geometrical test.
AP pointed out that, while tests using $d_A(z)$ or $d_L(z)$
can be affected by evolution in the sizes of ``standard rulers''
or the luminosities of ``standard candles,'' one can measure the
ratio of redshift separation distance to angular separation distance
assuming only that the structures under investigation are isotropic.
Recent implementations of this idea consider, instead of the
idealized spherical clusters discussed by AP, the statistical pattern
of clustering traced by quasars \citep{phillipps94,ballinger96,
matsubara96,popowski98,outram01,calvao01}, 
galaxies \citep{ryden95,nakamura98,nair99,matsubara01},
or the \lya\ forest \citep{hui99,mcdonald99,mcdonald01}.
Adopting the notation of \cite{phillipps94} and \cite{popowski98},
we define
\be
\label{h1}
h={\Delta z \over z \Delta \theta}
\ee
as the ratio of redshift separation to a ``redshift arc length''
for equal tangential and line-of-sight separations in physical coordinates,
assuming $\Delta z \ll z$.
For a fixed physical separation, $\Delta z$ is proportional to
$H(z)$ and $(\Delta\theta)^{-1}$ is proportional to $d_A(z)$, so
$h(z)$ is proportional to their product,
\be
\label{eqn:ap}
h(z)={1+z \over c z} H(z) d_{A}(z).
\ee
We will refer to $h(z)$ as the AP parameter.

What are the prospects for measuring these observables in the
next $5-10$ years?  Our remarks here will be qualitative and
somewhat speculative, but it is useful to approach the 
predictions of \S\ref{sec:results} with some sense of what may be achieved by 
different methods.

The prospects for distance measurements are the clearest and 
most well studied.  The rms scatter of the relation between peak
luminosity and light curve shape for SN Ia is only $\sim 10\%$
\citep{phillipps94,riess96}, so each well observed supernova allows a 
distance estimate with a $1\sigma$ statistical uncertainty $\sim 5\%$.
Current samples \citep{hamuy96,riess98,perlmutter99} are concentrated
at $z\sim 0$ and $z \sim 0.5-0.8$, but the approach can be
extended to $z\sim 1.7$ if the SNAP satellite
is built.  With samples of hundreds or even thousands
of supernovae, the statistical errors will become very small
indeed, and the measurement accuracy is likely to be limited
by systematic uncertainties such as dust extinction, possible
evolution of the progenitor population, and stability of photometric
calibration over a wide dynamic range.  A precision $\sim 1\%$ to
$z\sim 1.7$ seems plausibly achievable, and perhaps even
unduly pessimistic.

There are numerous other ways to measure the distance-redshift relation.
Measurements of the Sunyaev-Zel'dovich decrement and X-ray properties
of clusters can be combined to yield the angular diameter distance
\citep{birkinshaw99,molnar02}.  This method can provide an entirely independent
check on SN Ia results, and its limiting precision depends on the
size of well-observed cluster samples and the accuracy with which
internal cluster properties (particularly substructure) can be
understood.  The angular diameter distance can also be measured by
identifying a characteristic scale (such as the curvature scale of
the CDM power spectrum) in the angular clustering of distant clusters
\citep{cooray01},
or even by using the amplitude of this angular clustering in comparison
to theoretical predictions. 
\cite{roukema00} have already
applied a similar approach to a putative feature at $\sim 130\hmpc$ 
in the quasar power spectrum.  Galaxy counts 
in a deep redshift survey depend on the volume element
$dV \propto d_A^2(z) H^{-1}(z)$, providing yet another way to 
constrain distances.  The chief uncertainty in this approach is
evolution of the galaxy population, but \cite{newman00,newman01}
argue that this can be controlled in the DEEP redshift survey 
by measurement of galaxy circular velocities, allowing useful
constraints on $w$.  

At redshifts $z \ga 2$ all of these methods become difficult, but the
\lya\ forest offers an alternative probe out to $z\sim 4$.
The predicted correlation
of flux along lines of sight to quasar pairs depends on $d_A(z)$, and
measurements of this correlation will improve as more close pairs
are discovered and studied.  Such estimates of $d_A(z)$ would be 
somewhat model dependent, but the statistics of flux along individual
lines of sight can provide detailed checks of the assumed model.
This method has not been investigated in any detail (though 
\citealt{mcdonald01} presents relevant results), so it is hard to
know what precision can be achieved, perhaps a few percent.
Strong gravitational lensing statistics also test the equation of state
through their dependence on distance at various redshifts \citep{cooray99},
and constraints can also be obtained by measuring the source
redshifts in well understood lens systems \citep{yamamoto01}.
Finally, the first acoustic peak in the CMB power spectrum gives a 
high-precision measurement of the angular diameter distance to the
surface of last scattering, at $z=z_r \approx 1100$.  The uncertainty
in this determination is associated with the uncertainty in the parameters
that determine the sound horizon at $z_r$, which are themselves constrained
by the CMB power spectrum and other cosmological observations.
It is again hard to know just what precision will be obtained on
$d_A(z_r)$ itself, but a percent or better seems plausible.

Out to redshift $z\sim 1$, the main observational probe of the growth
factor $D_1(z)$, at least at present, is the mass function of galaxy
clusters.  Because clusters are rare objects that form from $2-3\sigma$
excursions of the initial Gaussian fluctuation spectrum, their 
predicted abundance is sensitive to the normalization of that spectrum,
and thus to the product
$\sigma_8 D_1(z)$, where $\sigma_8$ is the rms fluctuation of matter
in spheres of radius $8\hmpc$ at $z=0$ (see, e.g., \citealt{bahcall97}).  
However, because the cluster
mass function is steep, abundances are also sensitive to the accuracy
and precision of mass determinations \citep{frenk90},
and the limitation on measurements
of $D_1$ is likely to be systematic rather than statistical.
The combination of X-ray, Sunyaev-Zel'dovich, and weak lensing approaches
should reduce these systematic uncertainties below current levels.
To guess what level of precision is achievable for $D_1$, we note that 
current ``$2\sigma$'' uncertainties in the fluctuation amplitude $\sigma_8$
(for specified $\Om$) are $\sim 10\%$ (e.g., \citealt{eke96}),
though independent estimates can differ by more than this amount
even when the input data are similar (see, e.g., \citealt{seljak01}).
Balancing the difficulties of working at higher redshift against
the anticipated large improvements in cluster data, it seems reasonable
to hope for $\sim 5\%$ precision in $D_1$ out to $z\sim 1$, 
possibly better.  Recent discussions of the potential of galaxy 
cluster surveys for constraining $w$ include \cite{haiman01}, 
\cite{newman01b}, and \cite{weller01b}.

Cosmic shear is another potential probe of $D_1(z)$, measuring the
amplitude of surface density fluctuations
(see \citealt{huterer01} for a discussion in the context of
$w$ constraints).
Recent measurements already
yield a constraint on $\sigma_8$ (for fixed $\Om$) that is competitive
with determinations from the cluster mass function, with remarkably
good agreement of independent estimates (see \citealt{maoli01} and
references therein).
Measurements of shear for samples
of foreground and background galaxies with different photometric
redshifts should allow $\sigma_8$ and $D_1(z)$ to be disentangled.
Weak lensing will, at the least, provide an independent check on
estimates of $D_1(z)$ from cluster masses, and the ambitious
surveys now underway may eventually yield significantly better
precision.  At $z>2$, the most promising route to $D_1(z)$ is the
flux power spectrum of the \lya\ forest, which is related to the
underlying matter power spectrum in a fairly straightforward way
\citep{croft98,croft99,croft01,mcdonald00,gnedin01,zaldarriaga01}.
Current uncertainties in the rms fluctuation amplitude are $\sim 15\%$,
with roughly equal statistical and systematic uncertainties.
The former will decrease with larger samples such as those from
the Sloan Digital Sky Survey \citep{york00}, while the latter will
decrease with improved determinations of the mean \lya\ flux decrement,
improved numerical simulations to calibrate the relation between the
flux and matter power spectra, and the use of other statistics
to test the assumptions that enter these simulations.  At this point,
it is not clear where systematic uncertainties will limit the precision
of mass fluctuation measurements from the \lya\ forest, but $5-10\%$
seems a reasonable guess.  The other redshift at which we can
expect to determine $D_1(z)$ is the redshift of recombination, from
comparing the amplitude of the CMB power spectrum to that of today's matter
power spectrum \citep{doran01b}.  
Here measurement precision will be high, and the
limiting factor is the degeneracy of the fluctuation amplitude
with other parameters that affect the level of CMB anisotropy.

We note in passing that the cluster abundance, cosmic shear amplitude,
and \lya\ flux power spectrum are not ``pure'' measurements
of $D_1(z)$, since the distance-redshift relation affects the first
two through volume factors and lensing geometry, respectively, and
the Hubble parameter affects the third because the power spectrum
is measured in $\mkms$ units at the observed redshift.  Similarly,
the angular diameter distance is needed to identify angular scales in the 
CMB with lengthscales at $z=0$.  However, given the direct dependence
of these quantities on the mass fluctuation amplitude, it makes sense
to describe them primarily as probes of the growth factor.  Going
from an amplitude of fluctuations at redshift $z$ to a value of 
$D_1(z)$ also requires accurate knowledge of the fluctuation amplitude
today (i.e., of $\sigma_8$), which we are implicitly assuming will
emerge from the tightening web of CMB, large scale structure, cluster,
and weak lensing constraints.  
The obtainable precision on $D_1$ may be higher than the precision 
in $\sigma_8$ itself in the case of a differential evolutionary
measurement, such as the cluster mass function.

Lower limits to the age of the universe can be obtained by modeling
the stellar populations of the oldest galaxies observed at a given
redshift.  This approach has been used to argue against $\Omega_m=1$
models, for which the age scales as $t(z)=t_0(1+z)^{-3/2}$;
even relative to open models, the addition of a cosmological constant 
makes it substantially easier to understand the red colors and high
stellar mass-to-light ratios of high-redshift ellipticals
(e.g., \citealt{peacock98,vandokkum98}).  
\cite{lima00} have investigated the usefulness of galaxy ages as a
constraint on $w$.  Given the uncertainties associated with 
population synthesis modeling and dust extinction, precision of
10\% or better in $t(z)$ at high $z$ would seem highly optimistic.
However, age constraints can provide an upper limit on $w$ that allows
a consistency check with other estimates.  Exploiting this limit
requires accurate knowledge of $H_0$, which sets the overall normalization
of timescales.

The most promising targets for the AP test are quasars
\citep{phillipps94,ballinger96,matsubara96,popowski98,outram01}, the
\lya\ forest towards quasar pairs \citep{hui99,mcdonald99,mcdonald01},
and galaxies in the Sloan or 2dF redshift surveys
\citep{ryden95,nakamura98,matsubara01}.
The \lya\ forest approach is elegant, but \cite{mcdonald01} shows
that $h(z)$ at $z\sim 2-4$ is insensitive to $w$, and our results 
below reinforce this conclusion.  Instead, $h(z)$ at these redshifts
provides a good diagnostic of $\Om$ (and thus $\Op\approx 1-\Om$), 
with little dependence on $w$ if it is less than $-0.5$ \citep{mcdonald01}.
A precise value of $\Om$ is needed to get useful constraints on
$w$ with other tests, as we discuss and illustrate below.  Most studies
of the AP test with quasars or galaxies also focus on $\Om$ and 
$\Omega_{\Lambda,0}$ rather than $w$.  However, \cite{calvao01} have
examined constraints on $w$ that could be obtained with the 2dF 
quasar redshift survey, with encouraging conclusions.  They do not
present their results in the form of precision on $h(z)$, but
their projected sensitivity to $w$ must imply fairly good precision
at $z\sim 0.5-1$.

The Hubble parameter $H(z)$ is the observable most directly tied to
the Friedmann equation~(\ref{eqn:fr}).  One way to measure it is by
combining the volume-redshift or AP test with estimates of $d_A(z)$.
The \lya\ forest offers a more direct route because the width and
separation of features is determined largely by Hubble flow
\citep{weinberg97}.  Statistics like the threshold crossing frequency
are sensitive to the difference between open and flat CDM models
because of the difference in $H(z)$ \citep{weinberg99a}, and
measurements of the power spectrum shape can yield characteristic scales
in \kms\ units at the observed redshift, for comparison with scales
measured in $\hmpc$ at $z=0$ \citep{croft01}.  This method of measuring
$H(z)$ has not been investigated in any detail, so we do not know
what precision is attainable; it is likely to be set by the tradeoff
between $H(z)$ and other parameters that describe the temperature-density
relation of the diffuse intergalactic medium.  It is likely to work
better at $z\sim 2-4$ than at lower redshifts, where the observations
must be done from space and shock heated gas contributes more to 
the \lya\ forest \citep{dave99}, though even here the separation between
features might prove a useful diagnostic of the expansion rate.

At $z\la 1$, the skewness of the
cosmic shear distribution offers an alternative probe of $H(z)$.
\cite{hui99a} discusses the constraints on $w$ that can be obtained
by this method, which arise from the sensitivity of the predicted
skewness to the value of $\Omz$.  Since the
matter density is necessarily $\rho_m(z)=\Om \rho_{c,0} (1+z)^3$, 
the cosmology dependence of $\Omz$ comes from the critical density
$\rho_c(z)=3H^2(z)/8\pi G$, so in the context of our discussion it
makes sense to view weak lensing skewness as a measurement of $H(z)$.
Jimenez \& Loeb (2001) have proposed yet another route to measuring
$H(z)$, using the {\it relative} ages of galaxy populations at two
different redshifts (which can be determined more accurately than
the absolute ages, since some of the uncertainties in the population
synthesis models cancel out).  The ratio of redshift difference to 
age difference yields $dz/dt = -(1+z)H(z)$, where the equality uses
the definitions $(1+z)=a_0/a$ and $H = \dot{a}/a$.
Note that, while the \lya\ forest and weak lensing methods effectively
measure the ratio $H(z)/H_0$, the age difference method gives $H(z)$
in physical units.  Jimenez \& Loeb (2001) argue that percent-level
precision in $H(z)$ is achievable, in which case the uncertainty in
the ratio (which is the quantity sensitive to the equation of state)
is likely to be dominated by the uncertainty in $H_0$ itself.

There is significant degeneracy between the value of $\Om$ and the
value of $w$, since either lower $\Om$ or lower $w$ leads to greater
acceleration.  We assume that improving CMB and large scale structure
measurements will allow a precise determination of $\Om$ in the next few
years, independent of measurements of $d_A$, $D_1$, $t$, $H$, and $h$,
so that the power of these constraints can be brought to bear entirely
on the equation of state.  We will consider the impact of a 0.05
uncertainty in the value of $\Om$, and it is not obvious whether this
assumption is optimistic or pessimistic.  Apart from the determination
of $\Om$, the only role that we ascribe to the CMB is the measurement
of $d_A$ and $D_1$ at $z\approx 1100$.  It may be that CMB data 
can also yield constraints on the expansion history and $D_1$ at
lower redshifts, via the integrated Sachs-Wolfe (\citeyear{sachs67})
effect or lensing of anisotropies (see, e.g., \citealt{seljak96}),
but we do not know just what these constraints will be.
If the quintessence field is inhomogeneous, it will contribute to large-angle
CMB anisotropy \citep{caldwell98}, allowing
a probe of the dark energy independent of
the ones considered here, which are all based on the expansion history.

By focusing on this specific set of observables, we do not wish to imply
that this is necessarily the ground on which theory and observation
will be compared.  Presumably the constraints on the equation of state
from, say, weak lensing will be derived in terms of the weak lensing
observables themselves, without first extracting constraints on $D_1$,
$d_A$, and $H$ at various redshifts.  However, in trying to understand
the potential power of combining different observational approaches,
it is helpful to think in terms of the fundamental quantities that
they can measure.  In particular, two models that predict indistinguishable
results for $d_A(z)$, $D_1(z)$, $t(z)$, and $H(z)$ cannot be
discriminated by any combination of observations that depend only 
on these quantities.  Our focus on fundamental observables is also a helpful
way of estimating the level of precision needed for some observational
strategy to make a useful contribution to constraining the equation
of state and its history.

\section{Dependence of the Observables on the Equation of State}
\label{sec:results}

\begin{figure}[tbp]
   \label{Fig:GrandPlot}
   \centering
   \includegraphics[height=7.25in]{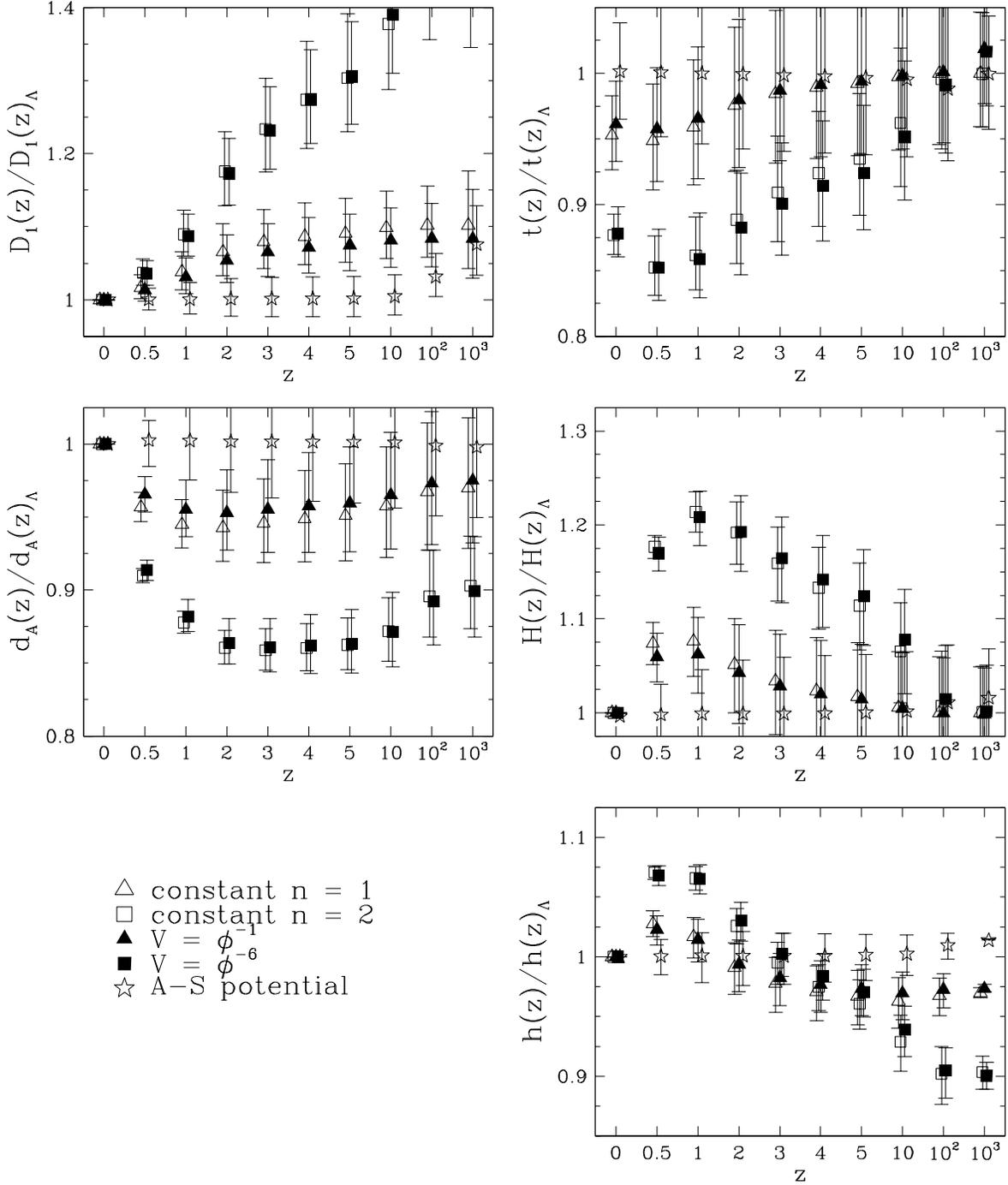}
   \caption{$D_1(z)$, $t(z)$, $d_A(z)$, $H(z)$, and $h(z)$ versus $z$ for the
   constant $n=1,2$ models, the fixed potential models $V=\phi^{-1}, \phi^{-6}$,
   and the Albrecht-Skordis potential.  
   The central value represents $\Om=0.4$ and the
   error bars are at $\Om=0.35, 0.45$.  Quantities are normalized to the
   $\Lambda$ model with $\Om=0.4$.  A small horizontal offset has been
   added to the points to allow them to be distinguished.
   The $\Om=0.45$ end of the error bar is usually the end further
   from a ratio of unity (or, for the A-S model, the end closer to the
   $n=1$ points), except for $D_1(z)$ and the high redshift ($h<1$)
   regime of $h(z)$, where the $\Om=0.45$ end is closer to unity.
   }
\end{figure}

Figure 3 presents our basic results for the 
quintessence models discussed in \S\ref{sec:quintessence}.  Each
panel shows the evolution of one of the five observables,
$d_A$, $D_1$, $t$, $H$, or $h$, out to redshift $z=10^3\approx z_r$.
Open triangles and squares represent constant-$w$ models with
$n=1$ and $n=2$ ($w=-2/3,\; -1/3$), respectively.
Filled triangles and squares represent $V=\phi^{-1}$ and $V=\phi^{-6}$
models, and stars represent the Albrecht-Skordis model with
the parameters stated in \S\ref{sec:quintessence}.  In all cases,
the points are computed assuming a flat universe,
$\Om=0.4$, $\Or=9.8\times 10^{-5}$, and $\Op=1-\Om-\Or$.
Furthermore, we normalize the value of each observable to the
value predicted by a pure-$\Lambda$ model ($n=0$, $w=-1$) 
at the corresponding redshift.  Thus, for example, the open triangle
at $z=0.5$, $d_A(z)/d_A(z)_\Lambda = 0.96$ implies that a precision
of 4\% (at the desired confidence level) is sufficient to distinguish
an $n=1$ model from a $\Lambda$-model using the angular diameter
distance at $z=0.5$, {\it if} $\Om$ is known perfectly. 
The ends of the error bars on each point show results for models
with the same equation of state but $\Om=0.35$ and 0.45, to illustrate
the impact of uncertainty in $\Om$.  (Ratios are still computed relative
to an $\Om=0.4$ $\Lambda$-model.) If the error bar on an 
observable overlaps a ratio of 1.0, then even a perfect measurement
of that observable at that redshift will not distinguish the model
from a $\Lambda$-model unless $\Om$ is known to better than 0.05.

For a given precision, the sensitivity of different observables peaks
at different redshifts.  The Hubble parameter sensitivity peaks at
$z\sim 1-2$, when the ratio of $\Opz$ values in different models
is large and the quintessence energy density is still large enough
to be dynamically important (see Figure~\ref{Fig:Omega_vs_z};
roughly speaking, it is the absolute difference in $\Omega_\phi$
in this Figure that matters for differences in $H$).
The sensitivity of $d_A(z)$ remains fairly flat, since even at
high $z$ the distance ``remembers'' the behavior of $H(z)$ at low
redshifts (see eq.~\ref{eqn:da}).  The age of the universe, by 
constrast, depends only on the Hubble parameter at redshifts 
{\it higher} than $z$ (eq.~\ref{eqn:age}), so the sensitivity
of $t(z)$ continues to increase almost all the way down to $z=0$.
Note that the sensitivity of an observable to the value of $n$,
displayed in Figure 3, may be quite different from the sensitivity
of that observable's $z-$derivative, which often peaks at lower
redshift.

The behavior of $h(z)$ is governed by the competing effects of
$H(z)$ and $d_A(z)$ --- from equations~(\ref{eqn:ap}) and~(\ref{eqn:da}),
one can see that $h(z)$ is proportional to the product of $H(z)$
and the average value of $H^{-1}(z)$ at lower redshifts.
At $z\approx 3$, $h(z)$ is very insensitive to $w$, as pointed
out by \cite{mcdonald01}, who emphasizes that this independence
of $w$ makes the AP test at this redshift an especially {\it good}
diagnostic of $\Om$.  The sensitivity to $w$ peaks at $z\sim 0.5$,
making quasar clustering better than the \lya\ forest as a probe
of the equation of state {\it per se}.  The Sloan survey's
luminous red galaxy sample \citep{eisenstein01} might also be
useful for this application \citep{matsubara01}.  
The sensitivity of $h(z)$ to $w$
grows again at $z \ga 5$, but the prospects for applying the AP
test at these redshifts seem very slim.

The sensitivity of $D_1(z)$ increases with increasing $z$, since
the growth factor depends only on clustering between redshift $z$
and redshift zero, and models with lower $n$ have larger 
$\Omz=1-\Opz-\Orz$ at all $z>0$.  For $n=1$, the sensitivity 
levels out at $z\ga 3$ as quintessence becomes dynamically
unimportant, but for $n=2$ the value of $\Opz$ is non-negligible
even at fairly high redshift (Figure~\ref{Fig:Omega_vs_z}).  

Figure 3 shows that $\sim 10\%$ measurements of any 
of these observables near their redshift of peak sensitivity can 
discriminate an $n=2$ model from a $\Lambda$-model, or from an 
$n=1$ model.  Since the expansion of an $n=2$ model with $\Om>0$ 
is always decelerating, it is not surprising that this model is
fairly easy to distinguish from a $\Lambda$-model with significant
acceleration at low redshift.  SN Ia measurements already rule out
this value of $n$ \citep{garnavich98}, and
Figure 3 implies that other observations are within
reach of confirming this result independently.  Distinguishing an $n=1$
model from a $\Lambda$-model is much harder, typically requiring
measurement precision of a few percent or better, and independent
precision on $\Om$ that is not much worse than the 0.05 represented
by our error bars.  Nonetheless, this level of discrimination is
clearly within reach of the improving SN Ia measurements, and the
discussion in \S\ref{sec:observables} suggests that several other
methods have a realistic hope of reaching the necessary precision on
a timescale of several years.  The sensitivity of our observables to $w$
would be slightly greater if we adopted $\Om = 0.3$ as our central value.
Furthermore, the precision that might be obtained from measurement of
these observables at multiple redshifts can be higher if the errors in the
separate measurements are uncorrelated; however, if the source of
uncertainty is systematic, it may produce correlated
errors at different redshifts.

Unfortunately, distinguishing any of the time-varying $w$ models that we have
considered from the closest constant-$w$ model looks all but impossible.
The $V=\phi^{-1}$ model tracks the $n=1$ model almost perfectly,
and the $V=\phi^{-6}$ model tracks the $n=2$ model with similar
faithfulness.  The close match of these models is unsurprising given
the plots of $n(z)$ and $\Opz$ in Figures 1
and 2.  The predicted differences between the power-law potential
models and the constant-$n$ models, and the redshift-dependence
of these differences, have the sign one would expect from 
Figure 1; the problem is simply that the time-dependence
of the equation of state predicted by these models is extremely weak.
Similarly, the Albrecht-Skordis model is virtually indistinguishable
from a pure-$\Lambda$ model because it has $n\approx 0$ at all
redshifts where quintessence is dynamically important, even though
it has a very different $n$ at $z>3$.  The one potential distinguishing
feature of the Albrecht-Skordis model is the value of $D_1$ at
recombination, which is about 7.5\% larger than that of a $\Lambda$-model.
This level of precision is plausibly within reach of future observations.
The distinguishability of the Albrecht-Skordis model would increase
if the equation-of-state transition were shifted towards lower redshift,
and vice versa.

Since we have so little empirical information about the nature of
dark energy, there is no reason to think that models presently in
the literature exhaust the possibilities for the time-dependence
of the equation of state.  We have therefore constructed a set of
``toy'' models that exhibit a wider range of behavior, so that
we can better understand the ability of observations to detect
time-variation if it is present.  For these models, we assume
that the redshift-dependence of $\rho_\phi$ is a broken power-law
of the form
\begin{equation}
\label{broken}
\rho_{\phi}(z) = {(1+z_c)^{n-m} (1+z)^{n} \over (1+z_c+z)^{n-m}}
\rho_\phi(z=0),
\end{equation} 
where $z_c$ is the critical redshift 
near which the scaling behavior of $\rho_\phi$  
changes from $(1+z)^m$ for $z \gg z_c$ to $(1+z)^n$ for $z \ll z_c$. The 
model therefore switches from an early time constant-$m$ case to a late 
time constant-$n$ case. This class of models allows us to
examine the effects of a more extreme change in the power-law
scaling of $\rho_\phi$ than is exhibited by the tracker models examined 
earlier.  (For the cases $m=n$ or $z_c=0$, this model reduces exactly to the
constant-$n$ models discussed earlier, while the $m=3$, $n=0$ case
resembles the Albrecht-Skordis model).  Note that a similar, but somewhat
different toy model was examined by Huterer and Turner (2001), who looked
at models in which $w$ was taken to be constant in discrete redshift
bins.

\begin{figure}[tbp]
   \centering
   \includegraphics{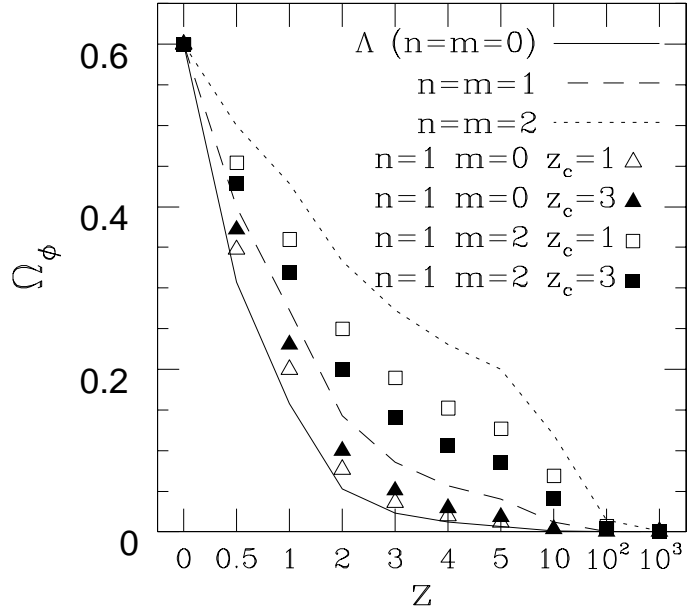}
   \caption{$\Omega_\phi$ as a function of redshift $z$ 
   for the indicated broken power-law models, defined by
   eq.~(\ref{broken}).
   The curves show three constant-$n$ models for reference.}
   \label{Fig:Omega_vs_z_BrokenModels}
\end{figure}

Figure \ref{Fig:Omega_vs_z_BrokenModels} shows $\Omega_\phi(z)$ for a variety
of broken power-law models.  The late time behavior ($n$) has
been fixed at $n=1$,
and early time 
behavior ($m$) has been taken to be $m=0$ or $m=2$, for two
different values of the critical redshift ($z_c = 1$ and $z_c=3$).
For comparison, we have also included three constant-$n$ models
($m=n$), namely $m=n=0,1,2$. 
As expected, $\Omega_\phi(z)$ in our broken power-law models deviates
from its behavior in the constant-$n$ models to a much greater extent than
is the case for the power-law potentials in Figure 2.

\begin{figure}[tbp]
   \centering
   \includegraphics[height=7.25in]{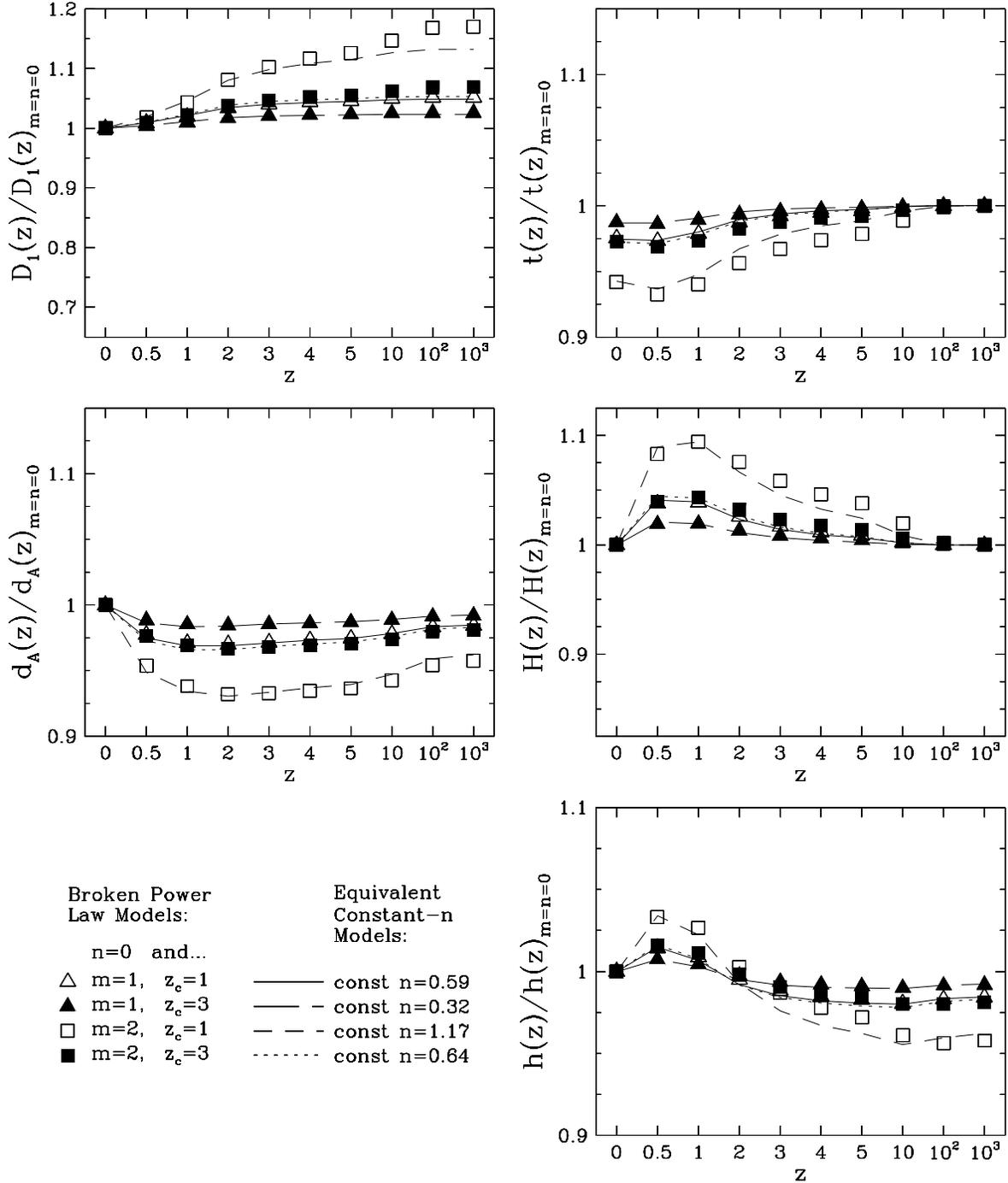}
   \caption{$D_1(z)$, $t(z)$, $d_A(z)$, $H(z)$, and $h(z)$
   versus $z$ for four broken power-law models, all having the
   same late-time behavior, $n=0$.  The points are normalized to
   the value of the given observable
   in the $n=0$ model at the same redshift.  For each set of points,
   a constant-$n$ model that has the same value of
   $H(z)$ at $z=1$ is shown.}
   \label{Fig:BrokenModels_n=0}
\end{figure}

\begin{figure}[tbp]
   \centering
   \includegraphics[height=7.25in]{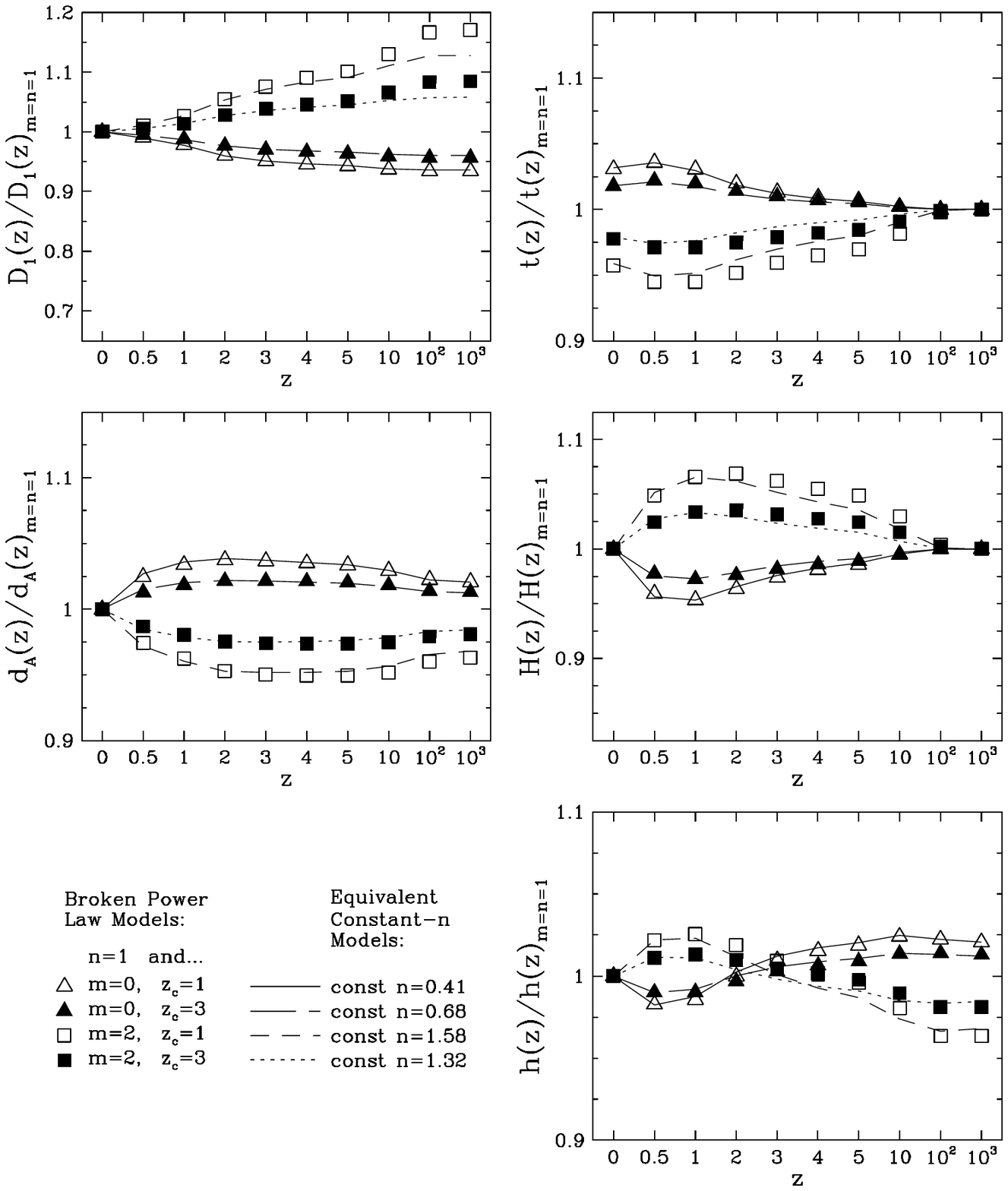}
   \caption{As Figure \ref{Fig:BrokenModels_n=0}, for
   the late-time behavior $n=1$.  The points are normalized
   to the $n=1$ model.
   }
   \label{Fig:BrokenModels_n=1}
\end{figure}

\begin{figure}[tbp]
   \centering
   \includegraphics[height=7.25in]{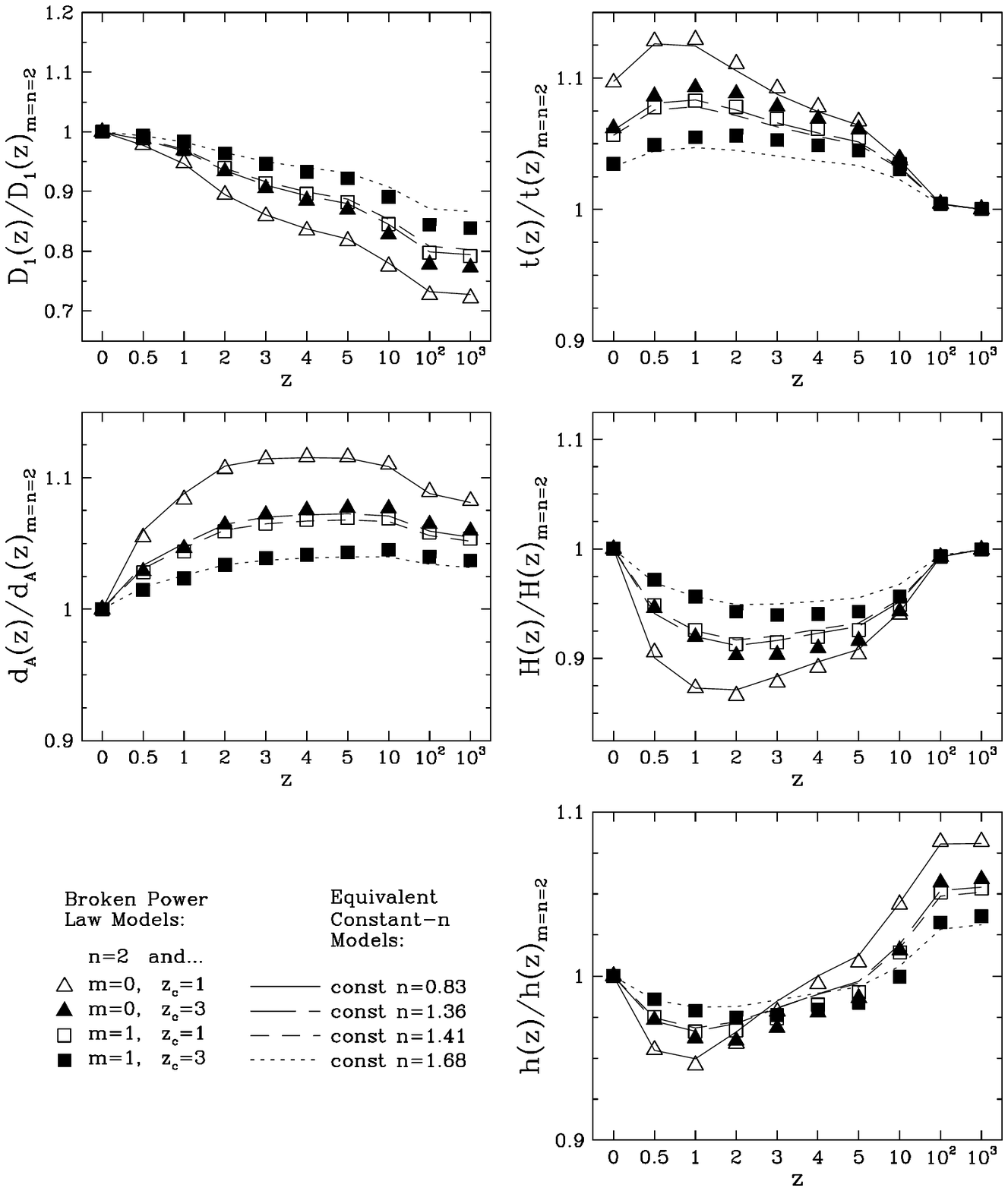}
   \caption{As Figure \ref{Fig:BrokenModels_n=0}, for
   the late-time behavior $n=2$.  The points are normalized
   to the $n=2$ model.  
   }
   \label{Fig:BrokenModels_n=2}
\end{figure}

In Figures \ref{Fig:BrokenModels_n=0}$-$\ref{Fig:BrokenModels_n=2} we 
examine our five observables
for the broken power-law cases $m,n=0,1,2$ and $z_c=1,3$.  Each page of
graphs shows models with a different late time behavior (different
value of $n$), and the observables on each page are normalized to the
corresponding constant-$n$ case:
$n=0$ in Figure 
\ref{Fig:BrokenModels_n=0}, $n=1$ in Figure \ref{Fig:BrokenModels_n=1}, 
and $n=2$ in Figure \ref{Fig:BrokenModels_n=2}. 
The deviation from a ratio of unity in each case shows the observational
effect of the 
break in scaling behavior.  This deviation is quite significant in many cases, 
often more than $10\%$ for $H$, $d_A$ or $t$, and up to almost 
$30\%$ for $D_1$. The broken power-law model 
is therefore clearly distinguishable from 
the constant-$n$ model that has the same
value of $n$ at $z=0$.

However, if we choose a constant-$n$ model that is matched to 
the ``average'' behavior of the broken power-law model instead of
the $z=0$ value of $n$, then this distinguishability vanishes.  
The curves in Figures \ref{Fig:BrokenModels_n=0}$-$\ref{Fig:BrokenModels_n=2} 
show, for each broken power-law case, the predictions of a
constant-$n$ model selected to produce the same value of the Hubble
parameter $H(z)$ at $z=1$. This matched constant-$n$ model
predicts nearly the same values for {\it all} observables
at {\it every} redshift as the corresponding
broken power-law model.  
By design, our toy model has a large change in $n$ (and thus $w$)
at low redshift, but this time-variation cannot be detected unless
the observables can be measured to extremely high precision.
Making the transition redshift $z_c$ higher or lower does not make the
time-dependence easier to detect; it just changes the effective 
average value of $n$, so the constant-$n$ model that matches the
broken power-law model is different from before.
For example, we have constructed models with $z_c = 0.5$, and
our results are quite similar, except that the corresponding
constant-$n$ model has a value of $n$ which is closer to the
low-redshift exponent in the broken power-law model.
A {\it sharper} transition would be easier to detect, but
our models already change the energy scaling index by
order unity over a redshift interval $\Delta z \sim z_c$
[and thus a time interval $\Delta t \sim H^{-1}(z)$], and a much
faster transition seems physically unlikely, although such
models can be constructed (see, e.g,
Weller and Albrecht 2001).

While our results suggest that generic variations of the equation
of state with redshift are essentially undetectable, there is one
exception to this rule that is worthy of note.  Our models with
high-redshift index $m=2$ predict values of $D_1(z)$ at $z=10^3$
that differ by a few percent from those of the constant-$n$ model
that matches the low-redshift observables.  This level of precision
might plausibly be achieved with comparisons of CMB anisotropy 
to local clustering, though substantial improvements in observational
data would be required.  [The values of $t(z)$, $H(z)$, and $h(z)$
also show percent-level deviations from the matched constant-$n$
model at $z\ga 3$, but we see no plausible routes to attaining the
necessary precision for these quantities.]
The behavior of the $m=2$ models is reminiscent of the Albrecht-Skordis
model, and the cause is similar: the energy density of the quintessence
component remains a non-negligible fraction of the critical density
out to fairly high redshift, so the gravitational growth of matter
clustering is correspondingly slower.

\section{Conclusions}
\label{sec:conclusions}

Our results both confirm and extend earlier work on this subject.
For any of the five observable quantities considered
here --- the angular-diameter distance $d_A(z)$, the Hubble
parameter $H(z)$, the age of the universe $t(z)$, the
linear growth factor $D_1(z)$, or the Alcock-Paczynski parameter
$H(z) d_A(z)$ ---
measurement with $\sim$10\% precision near the observable's
redshift of peak sensitivity would be sufficient to distinguish
an $n=2$ ($w = -1/3$) model from a pure cosmological constant, even if $\Om$
were known only to an accuracy of $\pm 0.05$.  Although this
value of $w$ is already ruled out by the SN Ia measurements,
our results suggest that other observations may soon be able to
independently confirm the result.  
Distinguishing an $n=1$ ($w = -2/3$) model from
a pure cosmological constant is much harder, demanding  
measurement precision of a few percent near the redshift of 
peak sensitivity, along with a determination of $\Om$
to within $\pm 0.05$.  Although this level of precision is currently
unavailable, it seems clearly within reach of improving SN Ia data,
and it is likely to be achieved by one or more of the other
observational methods discussed in \S\ref{sec:observables}.
Thus, while SN Ia surveys may provide the first precise determination of $w$,
a collection of other observations seems
likely to provide confirmation (or refutation!) 
of the measurement within a few years.

The sensitivity of the observables to the value of $n$ depends on
redshift in different ways, reflecting the links between these 
quantities and the expansion history.  The age $t(z)$ depends only 
on expansion at redshifts greater than $z$, so its sensitivity
to $n$ decreases monotonically with increasing $z$.
The linear growth factor, on the other hand, depends on clustering
from redshift $z$ to redshift zero, so the sensitivity of $D_1(z)$
increases monotonically with $z$.  The Hubble parameter $H(z)$
is most sensitive at $z\sim 1-2$, when $\Omega_\phi$ is substantially 
different from its present-day value but not so small that quintessence
is dynamically unimportant.  The sensitivity of the angular diameter
distance is fairly flat over a wide range of redshift.  The
sensitivity of the AP parameter is governed by competing effects of
$H(z)$ and $d_A(z)$, which cancel each other at $z\sim 3$.

Because of their different connections to the expansion history,
we hoped at the outset of this investigation that these observables
would provide complementary information about the history of the
equation of state, allowing a combination of measurements to detect
a time-variation of $w$ that could not be found by any one method
on its own.  Unfortunately, we find that the level of complementarity
is too weak to be useful in practice: models that make indistinguishable
predictions for one observable generally make indistinguishable 
predictions for all of them.
Of course, it is valuable to confirm an important result like a 
measurement of $w$ by independent methods, to check for systematic
errors or a breakdown of the assumptions implicit in each approach.
Also, different observables can provide complementary information
about $\Om$, precise knowledge of which is essential if one
hopes to constrain $w$.  However, once $\Om$ is known, the
constraints on the equation of state and its history will be dominated
by the single highest precision measurement; adding lower precision
measurements of other observables will give little additional purchase.

We find, furthermore, that none of the observables
holds much promise for distinguishing a quintessence model with a
time-dependent equation of state from an appropriately chosen
constant-$n$ model, even if one is highly optimistic about the
achievable precision and assumes perfect, independent knowledge of $\Om$.
Tracker models with $V(\phi)\propto \phi^{-1}$ and $V(\phi)\propto \phi^{-6}$
are effectively identical to models with constant $n=1$ and $n=2$,
respectively.  Models with an Albrecht-Skordis potential cannot
be distinguished from a pure-$\Lambda$ model, except, perhaps, by
a measurement of the growth factor at recombination from 
CMB anisotropy (a point we will return to shortly).
The fundamental difficulty is that, in any observationally viable model,
quintessence becomes dynamically important only at low redshift, so it 
affords little purchase for measuring redshift dependence of its 
equation of state.  Furthermore, as Figures
\ref{Fig:BrokenModels_n=0}$-$\ref{Fig:BrokenModels_n=2} demonstrate,
even a substantial transition in $n$ at low redshifts is very 
difficult to detect, since the value of $n$ at $z=0$ is not known
a priori, and time-variation must therefore be judged relative
to the constant-$n$ model that best mimics the time-variable model.
Our broken power-law models have substantial low redshift transitions
by design, but there is usually a constant-$n$ model that predicts
the same values of all observables to better than 1\% at all
observationally accessible redshifts.  We conclude that detecting
time-variation of the equation of state requires truly extraordinary
precision, unless this variation occurs on a timescale much shorter than
the Hubble time, which is possible but seems physically unlikely.
Sub-percent precision may be achievable by some methods (SN Ia observations
look to us like the best hope), but it requires controlling systematic
uncertainties, especially those that are correlated
among different redshift bins, very tightly.

Our conclusions in this regard agree with those of \cite{maor01},
who found that accurate measurements of the 
luminosity distance alone would be insufficient to determine the 
form of $w(z)$ for the dark matter energy component.  
\cite{wang01a} and \cite{tegmark01}
showed that SN Ia measurements
should be able to detect time-variation in the energy density $\rho_\phi(z)$,
but this only means demonstrating that $n>0$ ($w>-1$);
we agree that a significant departure from $n=0$ should be detectable,
but detecting time-variation of $n$ is far more challenging.
In a similar vein, despite fairly optimistic assumptions about the
prospects for the SNAP satellite, \cite{hutererturner01} find that error 
bars on the time-derivative of $w$ are quite large, and degrade
considerably with uncertainty in $\Omega_m$.
\cite{yamamoto01}
suggest that the form of the dark energy equation of state might be
determined by studying strong gravitational lensing systems, but their
results indicate that detecting time-variation is possible only with
extremely high precision measurements of the lensing systems, and then
only if $\Om$ is known precisely.
The principal significance of our results, relative to these earlier
papers, is that they apply to {\it all} proposed observable tests
based on the cosmic expansion history, since these tests always measure
some combination of $H(z)$, $d_A(z)$, $t(z)$, or $D_1(z)$.

Our investigation shows that there is one generic form of time-variation
in the equation of state that might be observationally detectable.
Constant-$n$ models with $n\geq 2$ ($w \geq -1/3$) are ruled out by
current data, but a time-variable model could have $n \geq 2$ at 
high redshift and a transition to low $n$ at low redshift when
quintessence becomes the dominant energy component.
The Albrecht-Skordis model displays just this behavior, since
the quintessence roughly tracks the matter energy density ($n\approx 3$)
along the exponential part of $V(\phi)$ but changes its equation of
state (to $w\approx -1$, $n\approx 0$) when it reaches the potential
minimum.  If $n\geq 2$ down to some fairly low redshift, then
the dynamical effects of quintessence are non-negligible (though small)
over a fair fraction of the post-recombination expansion history,
and they slow the progress of matter clustering.  The result is a
slight (few percent) mismatch between the value of $D_1$ at 
$z=z_r\approx 1100$ and the value expected for a constant-$n$ model
that matches the low redshift data; in observational terms, the level
of CMB anisotropy would be a few percent higher than anticipated.
\cite{doran01b} emphasize a similar point and discuss the relation
between CMB anisotropy and $\sigma_8$ in detail.
Detecting even this type of time-variation will be very challenging,
requiring a precise determination of the effective low-redshift
value of $n$, precise determinations of the present-day amplitude
of matter clustering and $\Om$, and the demonstration that any
excess CMB anisotropy does not arise from other sources, such
as tensor fluctuations, secondary anisotropies, or contaminating foregrounds.

The discovery of dark energy is an extraordinary cosmological achievement,
one that could happen only in the era of ``precision cosmology.''
If the equation of state of this dark energy is substantially different
from $p=-\rho$, or if it has been different in the recent past, then
that departure should be detected independently by several of the
ambitious observational efforts currently planned or underway.
A precise ($\sim \pm 0.1$) measurement of the low-redshift value of
$w$ would be another extraordinary achievement, ruling out many models
for the origin of dark energy and tightening the parameter space of others.
However, the information provided by different observable probes of
the cosmic expansion history, or by the same probe at different redshifts,
is mostly redundant rather than complementary, once $\Om$ has been
determined to high precision.
As a result, the next
step of detecting time-variation in the cosmic equation of state is likely
to prove extremely difficult.  If we are lucky, then the dark energy
has the kind of dynamical significance at high redshift or sudden
transition at low redshift that produces an observationally accessible
signature, though reading that signature will still require a 
combination of several cosmological measurements of unprecedented precision.
If we are not so fortunate, then the observable effects of the dark
energy will, for the foreseeable future, provide only two numbers with
which to describe it, the current energy density and an effective
low-redshift value of $w$ (or some equivalent pair of parameters).
Until a physical model comes along that accounts for these two
numbers in a natural way without adjustable inputs, the true nature
of the dark energy component is likely to remain mysterious.

\vskip 0.1 in

{\bf Acknowledgments}

A.M.L. and R.J.S. were supported in part
by the DOE (DE-FG02-91ER40690).  D.H.W. was supported in part
by the NSF (AST-0098584).  D.H.W. acknowledges the hospitality
of the Institute for Advanced Study and financial support of the
Ambrose Monell Foundation during the final phases of this work.
We thank L. Amendola, R. Jimenez, E. Linder, and the anonymous 
referee for helpful comments
on the manuscript.

\end{document}